# Photodissociation of $H_2$ on Ag and Au Nanoparticles: Effect of Size, and Plasmon versus Inter-Band Transitions on Threshold Intensities for Dissociation

Sajal Kumar Giri and George C. Schatz[*]

Department of Chemistry, Northwestern University, 2145 Sheridan Road, Evanston, Illinois 60208-3113, United States

E-mail: g-schatz@northwestern.edu

This paper provides new insights concerning the simulation of plasmon driven chemical reactions using real-time TDDFT based on the tight-binding electronic structure code DFTB+, with applications to the dissociation of $H_2$ on octahedral silver and gold nanoparticles with 19-489 atoms. A new component of these calculations involves sampling a 300 K canonical ensemble to determine the distribution of possible outcomes of the calculations, and with this approach we are able to determine the threshold for dissociation as a function of laser intensity, wavelength, and nanocluster size. We show that the threshold intensity varies as an inverse power of nanocluster size, which makes it possible to extrapolate the results to sizes that are more typical of experimental studies. The intensities obtained from this extrapolation are around a factor of 100 above powers used in the pulsed experiments. This is a closer comparison of theory and experiment than has been obtained in previous real-time simulations, and the remaining discrepancy can be understood in terms of electromagnetic hot spots that are associated with cluster formation. We also compare the influence of plasmon excitation versus inter-band excitation on reaction thresholds, revealing that for silver clusters plasmon excitation leads to lower thresholds, but for gold clusters, inter-band excitation is more effective. Our study also includes an analysis of charge transfer to and from the $H_2$ molecule, and a determination of orbital populations during and after the pulse, showing the correlation between metal excitations and the location of the antibonding level of $H_2$.

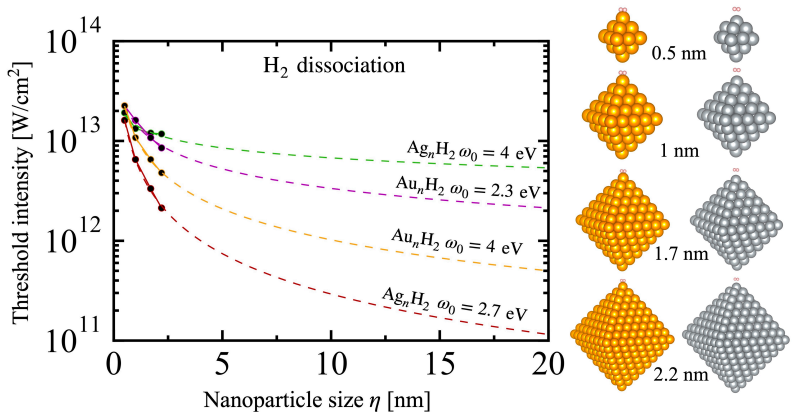

**Introduction**

Plasmon-driven chemistry (PDC) has recently gained substantial interest as it holds great promise for dramatically boosting the efficiency of sunlight absorption and solar energy conversion.[1-3] It is becoming a fast-growing field of surface photocatalysis where chemical reaction rates can be enhanced by several orders of magnitude, harvesting the electromagnetic field energy associated with localized surface plasmon resonance (LSPR) excitation of a metal nanoparticle (NP).[4-6] Within a few femtoseconds after plasmonic excitation, dephasing leads to highly excited electron-hole pairs, that subsequently can undergo scattering processes to become hot carriers that provide an elevated electronic temperature in the nanoparticle compared to the lattice temperature.[7,8] In addition, these hot carriers can be transferred to an electron (or hole)-accepting orbital in adsorbed molecules, leading to dissociation or other chemical reactions, either by directly populating antibonding electronic states, or by providing energy to overcome the barriers needed for dissociation, isomerization or other processes.[9,10] The electronic subsystem can also equilibrate with the lattice subsystem on the picosecond timescale. Another process involves direct photoexcitation of electron on the metal to the lowest unoccupied molecular orbital (LUMO) of the adsorbed molecules, which can also lead to dissociation dynamics. This can circumvent the problems associated with hot-carrier generation and decay but efficiency is limited by the requirement that the metal atoms be very near the nanoparticle surface, and that the energy of the charge transfer matches the excitation energy.[10,11] LSPR-induced dissociation of small important molecules including $H_2$ dissociation,[12-15] $H_2O$ splitting,[16,17] $CO_2$ reduction,[18,19] $NH_3$ decomposition,[20] and others,[21,22] has been reported. The efficiency of plasmon-enhanced dissociation is determined by several factors that influence hot-carrier generation and charge transfer dynamics, including the shape and size of NPs, and pulse properties like wavelength, pulse duration and intensity of the photoexcitation.[23,24]

The generation of hot carriers and charge transfer into surface molecules has been described using various theoretical models[25,26] and ab-initio calculations.[27,28] Real-time (RT) dynamics of NPs provide an explicit time-dependent perspective and a natural choice when a time-dependent electric field interacts with the system to generate hot carriers, in contrast to the more limited picture obtained from linear response frequency domain calculations.[29-31] The RT dynamics deals with the propagation of a single particle density matrix with an explicit time-dependent electric field. Some earlier studies have addressed RT-TDDFT simulations for NPs but mostly are restricted to small (few atom) systems or with serious approximations due to the systems' complexity and computational expense.[32-35] Schatz and co-workers[36] have demonstrated plasmon-mediated $H_2$ dissociation in a nanocavity performing RT-TDDFT calculations with a Gaussian pulse envelope but this required simplifications based on the jellium model. Dissociation was observed only for very high intensities (~$10^{14}$ W/cm$^2$) and extremely short pulse durations (1fs). Other theoretical works have reported the use of $10^{14} - 10^{15}$ W/cm$^2$ intensities and short pulse durations for dissociation to take place.[37-39] Such powers are close to the limit where relativistic electrodynamics would be required.[40] Pulsed laser-driven dissociation of molecules in presence of plasmonic NPs has been demonstrated experimentally with intensities and durations of 1-100 MW/cm$^2$ and 100-250 fs respectively.[41] It is of particular interest to explore the real-time dynamics of larger nanoparticles with sizes on the order of 10-20 nm and to establish the dependence of threshold intensity on particle size. Recently tight-binding density functional theory has been developed,[42] enabling the study of the RT dynamics of relatively large particles of Ag and Au atoms with dimensions up to 2-3 nm, but this is still far below the sizes studied in the experiments.

In this work, we investigate the dissociation of $H_2$ on Ag and Au nanoparticles induced by laser pulses using the RT-TDDFTB method implemented in the DFTB+ code. By introducing the canonical ensemble sampling of the initial vibrational coordinates and momenta of the molecules from an initial temperature of 300 K (but neglecting zero-point energies), we explore the effect of NPs size (from 0.5 nm to 2.2 nm), as well as driving frequencies, on the dissociation intensity threshold. This enables a determination of how the threshold intensity scales with particle size, providing a meaningful comparison with experimental intensity values for pulse durations in the range 100-250 fs. Two different driving frequencies are used to explore intra-band vs inter-band transition-assisted dissociations, and for each frequency, intensity, cluster size and element (Ag or Au) we perform trajectory simulations to determine dissociation probabilities. Furthermore, we have calculated orbital populations and time-dependent charge transfer between NPs and molecules along the trajectories to demonstrate how electron transfer drives the dissociation. The pulse amplitude vanishes after 25 fs for the pulse parameters we have chosen, but we propagate the density matrix up to 200 fs to observe the subsequent dynamics. Larger pulse durations might be more effective as more interaction time enables additional energy transfer, however, pulses longer than 200 fs for fixed fluence are not beneficial as plasmon decay processes (that are not fully described in DFTB+) convert electronic excitation into vibrational excitation.

The paper is organized as follows. In Section II we provide details of dynamical equations, pulses, calculations of spectra and nuclear trajectories. Then we discuss all the results in Section III and finally conclude in Section IV.

**Theory and Computation**

Under the real-time dynamics used in DFTB+, the single-particle density matrix $\rho(t)$ is evolved in time using an initial condition defined by the ground state Hamiltonian $H_0$, overlap matrix $S$, and density matrix $\rho_0$. The density matrix is propagated according to the Liouville-von Neumann equation within the Ehrenfest ansatz to include electron-nuclear dynamics[43] using

$$\frac{\partial}{\partial t}\rho(t) = -i(S^{-1}H\rho - \rho HS^{-1}) - (S^{-1}D\rho + \rho D^{\dagger}S^{-1}), \tag{1}$$

where $H$ is the Hamilton matrix, $S^{-1}$ is the inverse of the overlap matrix, and $D_{\mu\nu} = \langle \phi_\mu | \dot{\phi}_\nu \rangle$ are the non-adiabatic coupling matrix elements with localized atomic orbitals $|\phi\rangle$. Note that the derivative coupling term $D$ arises from nuclear motion, so in contrast to past work with frequency domain calculations for frozen structures to determine optical spectra, the RT approach leads rather naturally to dissociation of metal cluster and adsorbed molecule. The first term in the above equation is only for electron dynamics whereas the second term is responsible for coupling between electron and nuclear motions. To describe photoexcitation within this framework, the Hamilton matrix $H$ includes a time-dependent electric field through the interaction $V(t) = -\boldsymbol{\mu} \cdot \boldsymbol{E}(t)$, with an external electric field $\boldsymbol{E}(t)$ and transition dipole $\boldsymbol{\mu}$, where the latter is obtained summing over partial charges on the atoms. We note that both $D$ and $V(t)$ couple electronic states but in different ways. $D$ involves derivative couplings, while $V(t)$ includes the field-matter couplings. The force experienced by each atom is computed from the evolved non-equilibrium density matrix $\rho(t)$ and is determined by averaging over the distribution of electronic states:[31,42]

$$M\ddot{\boldsymbol{R}} = -\nabla_R [\text{Tr}(\rho(t)H(t))/\text{Tr}(\rho(t))], \tag{2}$$

where $\boldsymbol{R}$ denotes the atomic coordinates and $M$ specifies the masses of the atoms. Atomic charges are calculated using the Mulliken approximation[31,42], $q_A(t) = \text{Tr}_A[\rho(t)S]$, where $\text{Tr}_A$ indicates the trace over the orbitals centered on atom A.

We consider the following initial perturbation (delta function in time) in all possible spatial directions to calculate the real-time absorption spectra:

$$\boldsymbol{E}(t) = E_0 \delta(t) \boldsymbol{e}, \tag{3}$$

with the field amplitude $E_0$, and a unit vector $\boldsymbol{e} \in \{x, y, z\}$. In the frequency domain (the Fourier transform of $\boldsymbol{E}(t)$ in Eq.3), this perturbation corresponds to a constant value ($E_0$) independent of frequency, and therefore can include all possible frequency components in driving the system.

From the Fourier transform of the resultant time-dependent dipole along each spatial direction, we compute absorption spectrum as[44,45]

$$\sigma(\omega) = \frac{4\pi\omega}{3c} \Im[\text{Tr}[\boldsymbol{\alpha}(\omega)]], \tag{4}$$

where $c$ is the speed of light, and $\boldsymbol{\alpha}$ is the polarizability tensor, $\boldsymbol{\alpha}(\omega) = (\widetilde{\boldsymbol{\mu}}(\omega) - \boldsymbol{\mu}_0)/E_0$, with the frequency-dependent and initial (before the perturbation) dipoles $\widetilde{\boldsymbol{\mu}}(\omega)$ and $\boldsymbol{\mu}_0$ respectively. Alternatively, we compute spectra in the frequency domain using the linear response TDDFTB method through the calculation of oscillator strengths between the states. We use equivalent damping factors of the Lorentzian lineshape in the time and frequency domain to obtain the final smoothed spectra in order to compare spectra from both domains.

For laser pulse driven molecular dissociation dynamics, we have used an electric field of the following form

$$E(t) = E_0 e^{-(t-t_c)^2/T^2} \sin(\omega_0(t-t_c)), \tag{5}$$

with peak amplitude $E_0$, driving frequency $\omega_0$, peak position of the pulse $t_c$, and pulse duration $T$. During the calculation of nuclear trajectories, we consider different initial vibrational coordinates and momenta sampled from a canonical ensemble and calculate an ensemble of trajectories to generate a distribution of final outcomes. This type of calculation has not been considered in previous studies, but it is known to provide a more accurate treatment of the dynamics than the zero temperature approach involving a single trajectory, as has been used in the past. From the trajectories, we compute dissociation probabilities as a function of time by using a Gaussian distribution along inter-atomic distances to weight the reactive probabilities associated with each trajectory i.e., identifying the fraction of dissociative trajectories. A Gaussian function centered around 6 Å with a width of 1.3 Å is used, along with the assumption that as soon as the inter-atomic distance passes the center of the Gaussian (6 Å), the probability is kept at the peak value.

See the supporting information for the dependence of dissociation probability on Gaussian widths. We note that the explicit nuclear motion of all atoms has been included during the trajectory calculation.

We have performed all the electronic structure calculations and solved the real-time dynamics using the DFTB+ code.[42] A step size of 0.002 fs is used for the time evolution of the density matrix up to 200 fs. We have checked the propagation convergence using a smaller step size (smaller than 0.002 fs) and it turns out that a step size of 0.002 fs is reasonable to keep a balance between the propagation error and computational cost. For pulsed excitations, the laser pulse amplitude is chosen to vanish after 25 fs, but we further propagate the density matrix to observe the after-excitation dynamics, leading to charge transfer and dissociation of the molecules.

**Results and Discussion**

In this work, we have studied octahedral geometries of Ag and Au nanoparticles (NPs) of various numbers of atoms. We chose octahedrons for this work as this provides a simple motif for varying NP size over a wide range, and octahedron-shaped gold NPs have been studied experimentally,[46] having plasmon resonance properties that are similar (somewhat red-shifted) to what are found for spheres. Of course, the sharp points should provide hot spots with larger field enhancements than for spheres. Four different sizes have been used for each of the Ag and Au NPs with the number of atoms 19, 85, 231, and 489 corresponding to 0.5, 1, 1.7, and 2.2 nm dimensions respectively. Also, these geometries have been considered in some previous experimental[47-50] and theoretical[51,52] works. We have optimized each NP geometry using the DFTB method with hyb-0-2 parameters set for Ag NPs and auorg-1-1 parameters set for Au NPs.[53] A $H_2$ molecule is attached next to the tip of the optimized NP and then we further optimize the full system (NP with molecule). All the optimized geometries are shown in Fig.1 where the molecular axis is perpendicular to the z-axis. The inter-atomic distance of the hydrogen molecule is approximately 0.85 Å. We note that the gas phase equilibrium bond length of the molecule is approximately 0.74 Å but due to the adsorption on the NP surfaces here the bond length is slightly larger. The distance between the tip atom and molecule for systems involving Ag (Au) atoms is around 2.5 Å (1.8 Å). These optimized system parameters are comparable to previous DFT calculations for a tetrahedral $Au_6H_2$ system.[15]

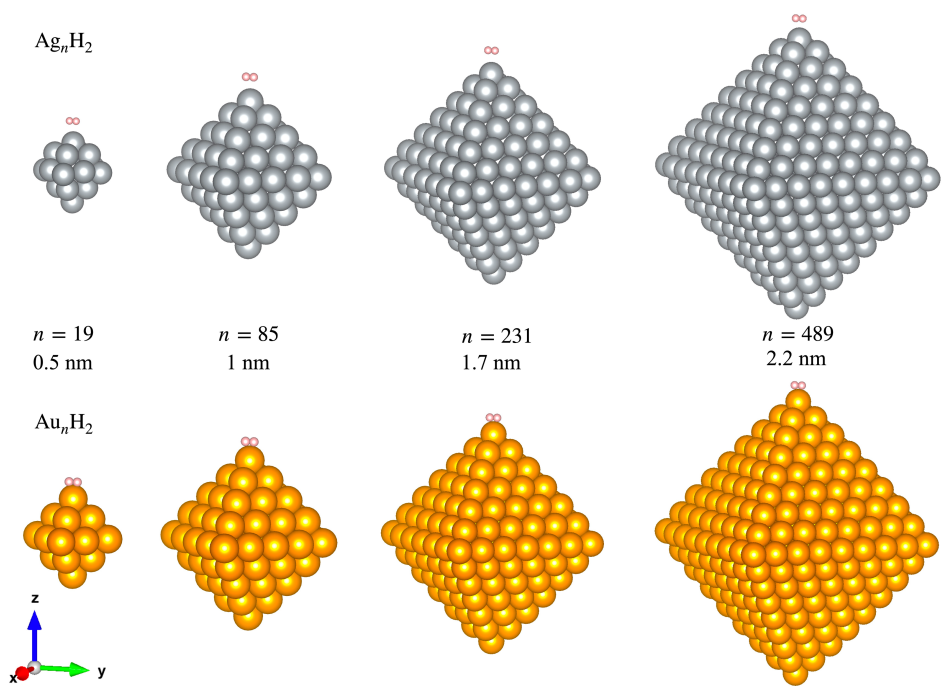

**Figure 1:** $H_2$ molecule adsorbed on octahedral Ag and Au nanoparticle surfaces of different dimensions determined by the number of atoms $n$, ($n = 19, 85, 231,$ and $489$). All the geometries are optimized with the DFTB method using the parameters set hyb-0-2 and auorg-1-1 for Ag and Au NPs respectively.

As a test of the RT code, we compute absorption spectra solving both the real-time dynamics (RT-TDDFTB) through the computation of time-dependent dipoles and linear response TDDFTB frequency domain calculations. In the linear response limit (at low intensities involving a one photon process), both spectra should match each other. For the time domain spectra, we solve the Liouville-von Neumann equation (Eq.1) with a delta kick perturbation described in Eq.3. From the resultant time-dependent dipole along each spatial direction, we calculate absorption spectra through Eq.4. We set the peak amplitude of the perturbation ($E_0$) to a small value 0.001 V/Å which corresponds to a peak intensity of $1.3 \times 10^7$ W/cm$^2$. We use the TDDFTB method in the Casida formulation in computing oscillator strengths between ground and excited states. Spectra from both methods agree very well with each other. Throughout we have used a damping factor of 0.16 eV in computing spectra. See the supporting information for the comparison between the time and frequency domain spectra for $Ag_{19}H_2$ and $Au_{19}H_2$ NPs.

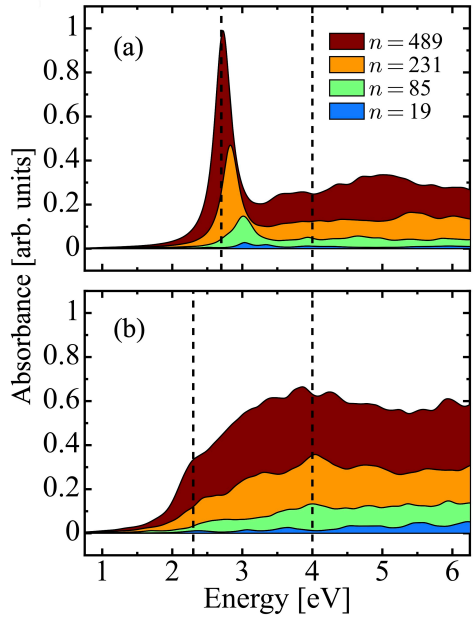

**Figure 2:** Real-time absorption spectra for (a) $Ag_nH_2$ and (b) $Au_nH_2$ systems. The two frequencies used for the dissociation dynamics are shown by vertical dashed lines.

Real-time absorption spectra for all systems are shown in Fig.2. The absorption intensity increases gradually with the size of the NPs as the density of states (DOS) increases. For Ag NPs, the LSPR peak appears between [2.5, 3] eV and shows a prominent red shift with increasing NPs size. The peak shifts towards lower energy by around 0.25 eV when we increase the number of atoms from 19 (0.5 nm) to 489 (2.2 nm). This trend was observed in previous studies of 0.5-2.0 nm clusters[54] where it was attributed to quantum size effects, and is distinct from the red shift that is seen for much larger octahedral NPs[49] which arises from electromagnetic depolarization effects.[55] The high energy flat signals (above 3 eV) originate from inter-band transitions i.e., from *d* to *sp* band transitions. For Au NPs, the plasmon peaks are relatively broader due to the short lifetime of the excited states compared to Ag NPs. As a result, it becomes difficult to distinguish the plasmon peaks from high-energy inter-band transitions for the smaller particles. The plasmon peak for the $Au_{489}H_2$ system arises at around 2.3 eV, which is similar to the plasmon in larger gold NPs with roughly spherical shape.[55,56] Here the plasmon peak is more pronounced and is distinct from the high-energy inter-band transitions. This also agrees with the previous study by Berdakin and co-workers[57] where plasmon peaks start to appear distinct from inter-band transitions for icosahedral Au NPs involving more than 300 atoms. We note that the broad oscillating structures in the Au NP spectra in Fig.2 at energies above 2.3 eV were also observed by Schatz and co-workers[58,59] for tetrahedral Au NPs using the TDDFT calculations.

We use two different driving frequencies $\omega_0$ = 2.7 eV (2.3 eV) and $\omega_0$ = 4 eV for Ag (Au) NPs to explore the dissociation dynamics of $H_2$ due to plasmon and inter-band transitions. These two frequencies are labelled in Fig.2 with vertical dashed lines and the width around each line is inversely proportional to the pulse duration $T$. The driving field described in Eq.4 is used in the simulation. We vary the laser peak intensity through the parameter $E_0$. We consider the *z*-axis as the laser polarization direction for all systems. Note that along this direction the tip atom of the cluster points towards the molecule and appears to be more suitable for bond dissociation compared to other possible directions. We have included an example pulse and its frequency content (Fourier transform) in the supporting information.

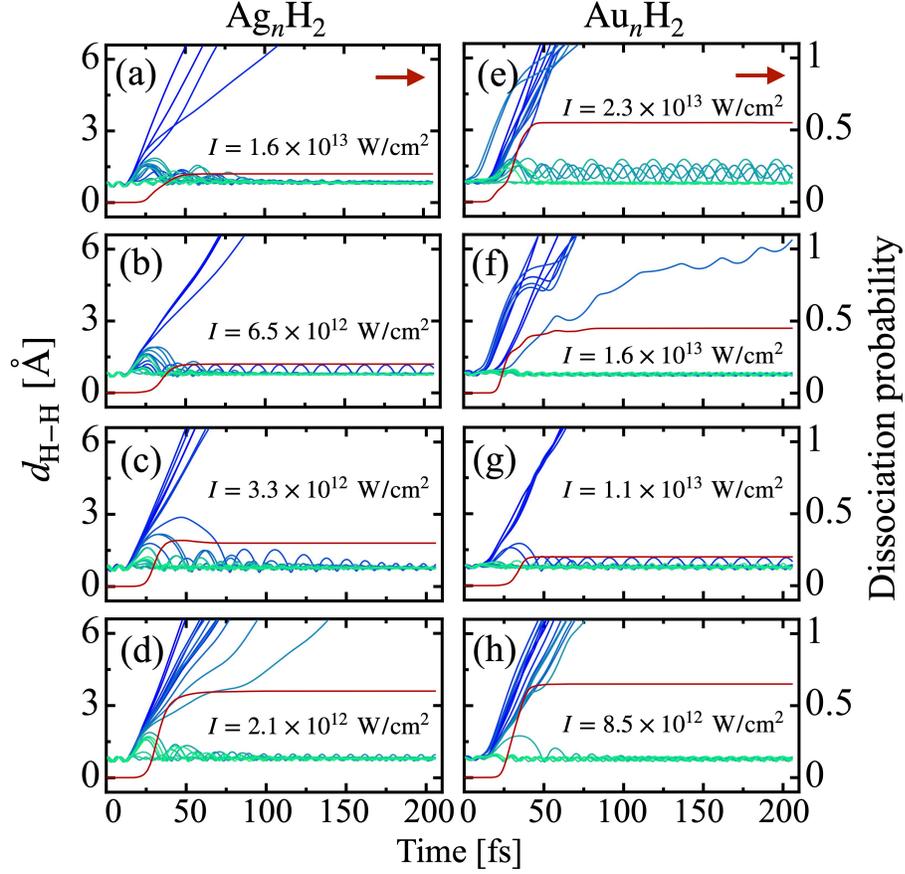

**Figure 3:** Trajectories showing inter-atomic distances and dissociation probabilities for (a-d) $Ag_nH_2$ and (e-h) $Au_nH_2$ systems with driving frequencies 2.7 eV and 2.3 eV, respectively, corresponding to plasmon-driven dissociation. From top to bottom, the size of the particle increases, $n = $ 19, 85, 231, and 489 respectively. Threshold intensities are given for all plots, denoting intensities where dissociation starts. The red lines are the ensemble-averaged dissociation probability (obtained through the Gaussian binning along bond distances) as a function of time. Here we have calculated 20 trajectories for each plot and sorted the color code from green to blue for non-dissociative and dissociative trajectories.

Laser pulse-driven dissociative trajectories for $Ag_nH_2$ ($Au_nH_2$) systems with $\omega_0 = 2.7$ eV (2.3 eV) are shown in Fig.3 where we have plotted inter-atomic distances $d_{H-H}$ for an ensemble of trajectories and dissociation probabilities (averaging the trajectories) as a function of time. Note that these frequencies correspond to plasmon-driven dynamics. We employ different pulse intensities for each plot, which are the lowest intensities necessary to initiate dissociation. We find that the dissociation probability increases linearly with the pulse intensity, from a threshold intensity up to a plateau where the probability is 1. The linear increase is because more hot carriers are generated with increasing photon flux and a one photon transition is sufficient to enable hot carrier generation leading to dissociation near threshold. The linear dependence was also observed in the previous experimental study of $H_2$ dissociation on a Au NP surface.[12,13] Most of the dissociations take place during interaction with the pulse i.e., within 25 fs. At low intensities, dissociation strongly depends on the initial vibrational phases, and therefore only a fraction of trajectories dissociate. For the remaining trajectories, the inter-atomic distance $d_{H-H}$ shows coherent oscillation around the equilibrium bond length. We note that for most of these non-dissociative trajectories, molecules desorb from the surface very quickly within 20-30 fs. At high intensities, the dissociation probability increases as shown in Figs.4(a, b) for the $n = 231$ systems with intensities chosen as the threshold intensity for the $n = 19$ systems. We found that beyond certain intensities, all the trajectories dissociate. The time-dependent dissociation probability calculated from an ensemble of trajectories through Gaussian binning along inter-atomic distances increases during the pulse interaction and saturates to a finite value within 40 fs. We note that there are interactions between dissociated atoms and the NPs as shown in Fig.5. Therefore, the dissociated outgoing trajectories are not completely smooth at the threshold intensities while at high intensities the dissociated atoms gain sufficient velocities to leave the surface as soon as the molecule dissociates, resulting in smoothed dissociative trajectories as shown in Figs.4(a, b).

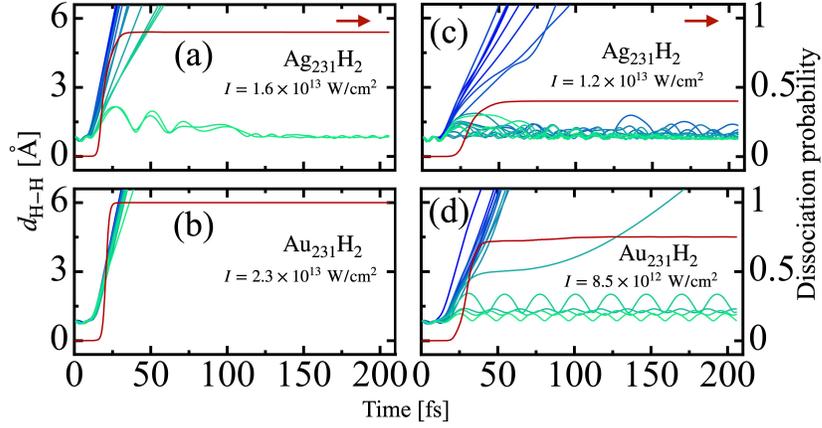

**Figure 4:** Same as Figs.3(c, g) but for (a, b) high intensities, and (c, d) $\omega_0 = 4$ eV.

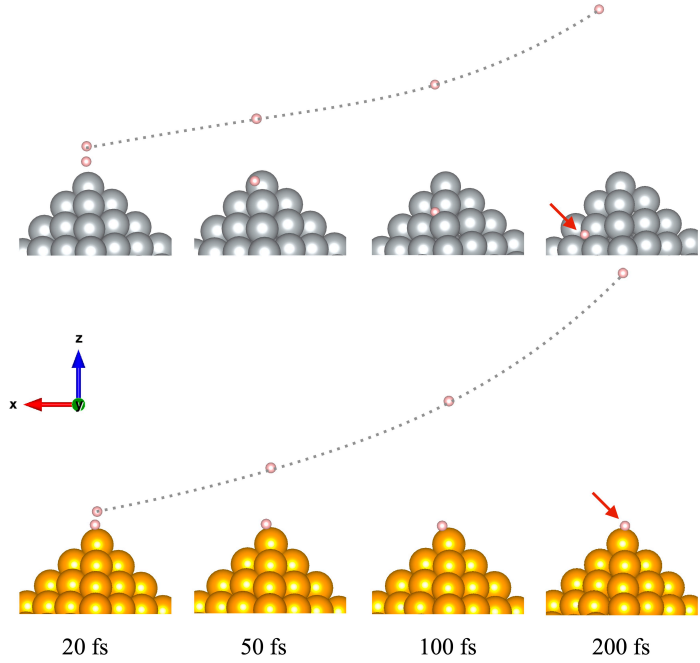

**Figure 5:** Geometries along the trajectories (a single trajectory for each system $Ag_{231}H_2$ or $Au_{120}H_2$) at 20, 50, 100, and 200 fs from Figs.3(c, g). Here $d_{H-H} = 23.3$ Å (22.8 Å) for Ag (Au) NPs at 200 fs.

Trajectories for the $Ag_{231}H_2$ and $Au_{231}H_2$ systems with $\omega_0 = 4$ eV are shown in Figs.4(c, d). High-frequency driving shows similar dissociation but involves different intensities than the corresponding low-frequency driving for the dissociation to take place. We note that the isolated molecule dissociates at even higher intensities than we have presented. This confirms that dissociation is enhanced by the presence of NPs and the enhancement factors strongly depend on the NP size and driving frequencies.

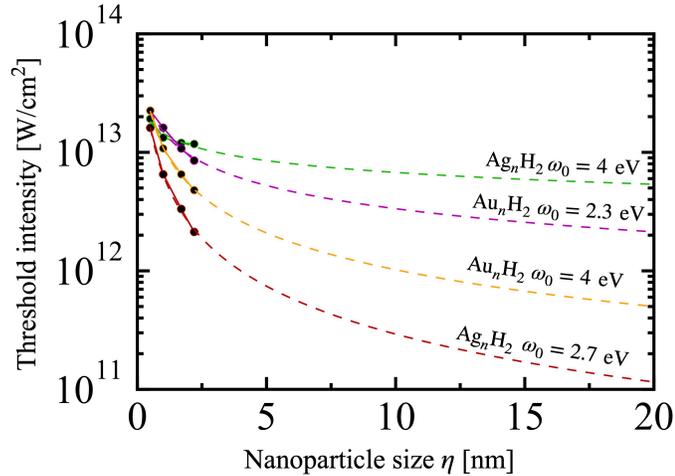

**Figure 6:** Threshold intensity ($I_t$) as a function of particle size $\eta$ for $Ag_nH_2$ and $Au_nH_2$ systems with two different driving frequencies.

**Table 1:** Fitting parameters, $a$ and $b$, for threshold intensity as a function of particle size, $I_t = a \times \eta^b$.

| System | $\omega_0$ [eV] | $a$ | $b$ |
|---|---|---|---|
| $Ag_nH_2$ | 2.7 | $6.4 \times 10^{12}$ | -1.34 |
| | 4 | $1.45 \times 10^{13}$ | -0.34 |
| $Au_nH_2$ | 2.3 | $1.5 \times 10^{13}$ | -0.65 |
| | 4 | $1 \times 10^{13}$ | -1.03 |

The above calculations are used to determine the threshold intensity for dissociation i.e., the minimum intensity required for dissociation to take place. Interestingly, the threshold intensity decreases with increasing particle size for both Ag and Au NPs but with different decay rates depending on the driving frequency as shown in Fig.6. We fit the decay curves with the functional form $I_t = a \times \eta^b$, where $I_t$ is the threshold intensity, $\eta$ is the particle size, and $a$ and $b$ are parameters to be optimized. The fitting parameters $a$ and $b$ are listed in Table 1. All the decays are approximated well with the same functional form $I_t = a \times \eta^b$. The Table shows that $b$ is negative and hence the threshold intensity decreases with particle size. The decay rate is maximum ($b = -1.34$) with minimum amplitude ($a = 6.4 \times 10^{12}$) for $Ag_nH_2$ systems with plasmon frequency ($\omega_0 = 2.7$ eV). This conclusion is in agreement with a recent experiment where Ag NPs are found to be more efficient than Au NPs in generating hot electrons.[24] On the other hand, for $Au_nH_2$ systems, the high frequency ($\omega_0 = 4$ eV) driving provides a larger decay rate compared to the low-frequency driving as there is a significant inter-band transition amplitude, see Fig.2b, while the plasmon amplitude is weak. For $Ag_nH_2$ particles there are dominant plasmonic transitions for $n \geq 85$ and the plasmon resonance shows larger efficiency for dissociation compared to the inter-band transition.

To provide insight about the comparison of calculated and experimental threshold intensities, we extrapolated the plots to large particle sizes up to 20 nm where the intensities are reduced significantly for large $|b|$ values. For $Ag_nH_2$ at 2.7 eV (with the maximum dissociation rate), the threshold intensity decreases from $10^{13}$ to $10^{11}$ W/cm² when we increase the particle size from 0.5 nm to 20 nm. We note that the peak intensity can be further lowered to $10^{10}$ W/cm² when we scale the pulse duration by a factor of 10 in order to compare with pulses of duration 200 fs used in the experiment.[41] This is still 2 orders of magnitude larger than the experimental peak amplitudes. However, this can be compensated by including aggregated particles in the study, as this would enable stronger hotspots, as known from surface-enhanced Raman spectroscopy (SERS) studies and electrodynamics theory,[60,61] and it would also provide a mechanism for trapping of the dissociating molecule in the hot spot, which would enhance charge transfer efficiency between nanoparticles and molecules.[36]

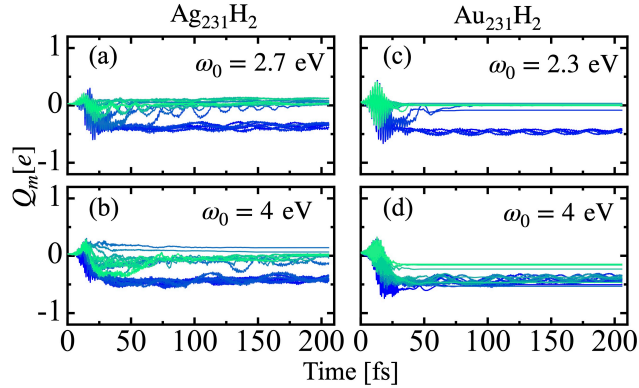

**Figure 7:** Trajectories for the molecular charge ($Q_m$) for $Ag_{231}H_2$ ($Au_{231}H_2$) with frequencies 2.7 (2.3) eV and 4 eV. Threshold intensities are used from Figs.3(c, g) for low frequencies and Figs.4(c, d) for high frequencies. The same color code is used for respective trajectories compared to Figs.3(c, g) and Figs.4(c, d) i.e., the color code from green to blue represents non-dissociative to dissociative trajectories.

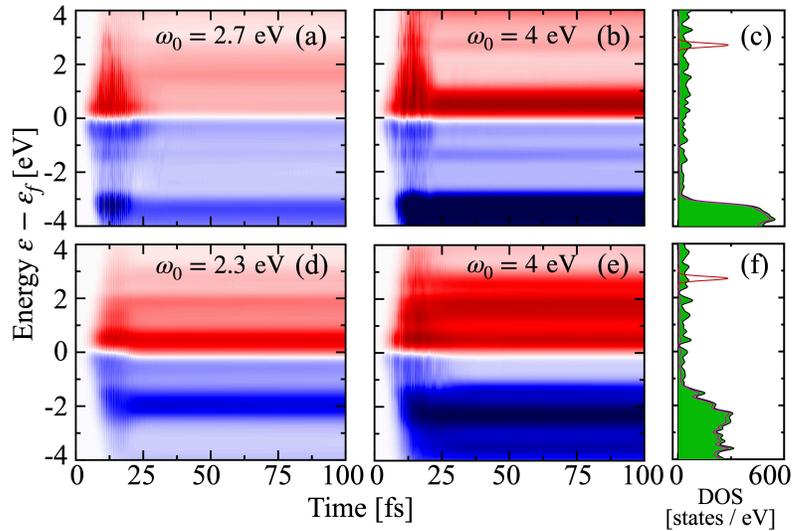

**Figure 8:** Orbital populations as a function of energy $\varepsilon - \varepsilon_f$ and time. The red (blue) color represents population gain (de-population) with positive (negative) amplitudes. The same (threshold) intensities are used as in Fig.7. (c, f) Density of states (DOS) where magenta lines represent $d$ band only and green shaded areas are for the complete systems (top) $Ag_{231}H_2$ and (bottom) $Au_{231}H_2$. Red lines (in the DOS panels) represent the antibonding orbital of $H_2$.

Figure 7 displays the time evolution of molecular charge $Q_m$ i.e., the combined charge of the two hydrogen atoms as obtained through the Mulliken approximation for the $Ag_{231}H_2$ and $Au_{231}H_2$ systems with the threshold intensities. Initially, the molecule is slightly positive with $Q_m \approx 0.05\ e$ when it is adsorbed on the surface of a NP. After the pulsed excitation, there is an exchange of electrons between two sub-systems and $Q_m$ follows the laser frequency oscillations. We note that the total charge is conserved at all times i.e., $Q_{NP} + Q_m = 0$, where $Q_{NP}$ is the NP charge. There is an effective transfer of electrons to the molecule at a certain time after the pulse is turned on (near the peak intensity of the pulses) and the average charge on the molecule becomes negative, corresponding to the net transfer of electrons. Similar to inter-atomic distance trajectories, the electron transfer process also depends on the initial vibrational phases. It is interesting to notice that only dissociative trajectories exhibit substantial electron transfer (the same color code is used as in Figs.3 and 4). This confirms that dissociation is driven by the electron transfer process. After bond dissociation, one of the dissociated hydrogen atoms sticks to the surface of the NPs (as shown in Fig.5), resulting in negative partial charges on the dissociated atoms at longer times. Electron back transfer from the molecule to NP is also observed for some trajectories when the molecule or dissociated atoms leave the surface of the NP and the final charge at longer times becomes nearly zero. In the earlier RT-TDDFT work,[36] similar charge transfer dynamics of $H_2$ in Au NPs were also observed.

Time-dependent orbital populations for the $n = 231$ systems with plasmonic and inter-band transition frequencies are shown in Fig.8. Here we show population changes compared to the initial populations (i.e., the populations prior to the pulse interaction). Due to the interaction with pulses, there is a transfer of populations across the Fermi level (with energy $\varepsilon_f$) from lower energy to higher energy states covered by the bandwidth of the pulses in the frequency domain. As expected plasmon excitations populate low energy states near the Fermi level while the inter-band transitions are extended to high energy states with dominant *d* to *sp* transitions. In comparison to Ag, Au NPs have a broader energy range for the *d* band that is closer to the Fermi level, which allows significant inter-band transitions to populate states above the Fermi level. All the populations show coherent oscillations following the laser frequencies during the pulse, and then after the pulse, the coherence de-phases very quickly. These oscillations are known to appear upon laser excitation of NPs which are referred to as "sloshing" and "inversion".[17,62,63] During the pulse interaction (within 25 fs), adequate populations accumulate near the antibonding orbital of the $H_2$ molecule (the red curves in Figs. 8c and 8f) which weakens the molecular bond and eventually the molecule dissociates.

**Conclusions**

This work demonstrates the laser pulse driven dissociation of $H_2$ on octahedral Ag and Au nanoparticle surfaces based on real-time density functional tight binding calculations within the canonical ensemble trajectory approach. We have explored the effect of nanoparticle size and driving frequencies on the dissociation intensity threshold, and we have also characterized the influence of pulse width. Calculations are performed for nanoparticles with varying numbers of atoms up to 489. As expected, a plasmon peak appears for Au nanoparticles only for a large number of atoms (more than 231) whereas Ag nanoparticles show a plasmon peak for much smaller particle sizes. In order to investigate plasmon vs inter-band assisted dissociation of molecules, two different driving frequencies are used in the real-time dynamics. Both particle size and driving frequency are found to be crucial in driving dissociation dynamics and hence the intensity threshold. Interestingly, we found that the intensity threshold decreases with particle size following a similar functional form irrespective of driving frequencies and systems. Using the functional fittings, we scale the results to larger nanoparticle sizes up to 20 nm, for which the largest decay rate is obtained for Ag nanoparticles excited at the plasmon frequency. In this case, the threshold power is about a factor of 100 larger than has been used in pulsed experiments, which is a result that is consistent with known enhancements that arise when clusters of particles rather than isolated particles are considered. For Au nanoparticles, inter-band transitions are more effective than plasmon excitation, a result that has also been seen in experiments,[64] and the threshold power is similar to that for Ag. Additionally, we have calculated the charge on the molecule as a function of time along each trajectory. After the pulsed excitation, we observe electron transfer from the nanoparticle to the molecule which initiates the dissociation process. We have also analyzed these results in terms of the orbital populations as a function of time.

**ASSOCIATED CONTENT**

The Supporting Information is available free of charge at

**AUTHOR INFORMATION**

Corresponding Author

George C. Schatz – Department of Chemistry, Northwestern University, 2145 Sheridan Road, Evanston, Illinois 60208, United States;

Email: g-schatz@northwestern.edu

Author

Sajal Kumar Giri – Department of Chemistry, Northwestern University, 2145 Sheridan Road, Evanston, Illinois 60208, United States

ACKNOWLEDGMENT

This research was supported by the Department of Energy, Office of Basic Energy Sciences through grant DE-SC0004752.

62. Townsend, E.; Bryant, G. W. Which resonances in small metallic nanoparticles are plasmonic? *J. Opt.* **2014**, 16, 114022.

63. Ma, J.; Wang, Z.; Wang, L. W. Interplay between plasmon and single-particle excitations in a metal nanocluster. *Nat. Commun.* **2015**, 6, 10107.

64. Wu, Y.; Yang, M.; Ueltschi, T. W.; Mosquera, M. A.; Chen, Z.; Schatz, G. C.; Van Duyne, R. P. SERS study of the mechanism of plasmon-driven hot electron transfer between gold nanoparticles and PCBM. *J. Phys. Chem. C* **2019**, 123, 29908-29915.

# Supporting Information:

# Photodissociation of $H_2$ on Ag and Au Nanoparticles: Effect of Size, and Plasmon versus Inter-Band Transitions on Threshold Intensities for Dissociation

Sajal Kumar Giri and George C. Schatz[*]

Department of Chemistry, Northwestern University, 2145 Sheridan Road, Evanston, Illinois 60208, United States

E-mail: g-schatz@northwestern.edu

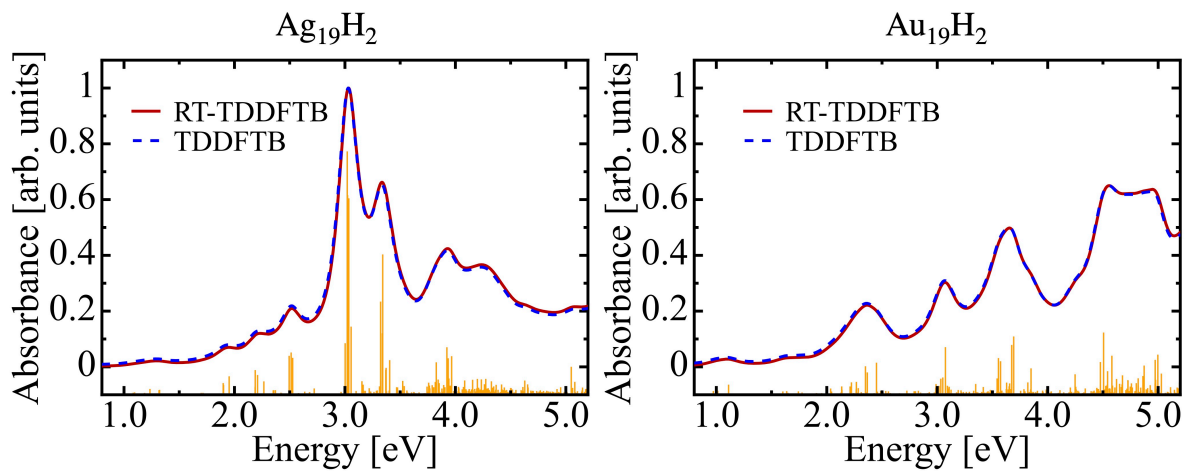

**Figure S1:** Comparison of absorption spectra obtained through RT-TDDFTB (time domain) and TDDFTB (frequency domain) methods. Vertical orange lines represent oscillator strengths.

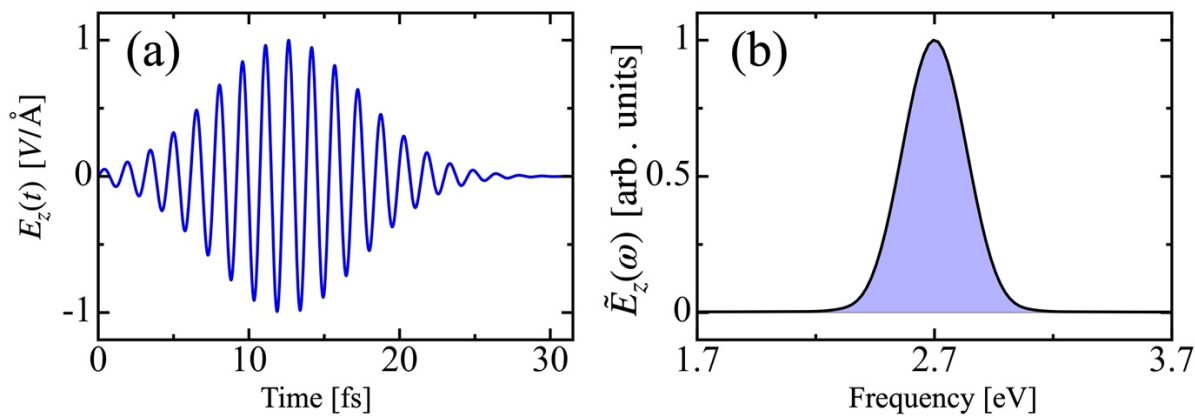

**Figure S2:** (a) Time dependent field $E_z(t)$ and (b) its frequency content (i.e., the Fourier transform of $E_z(t)$) with polarization along z-axis. Here, $\omega_0 = 2.7$ eV.

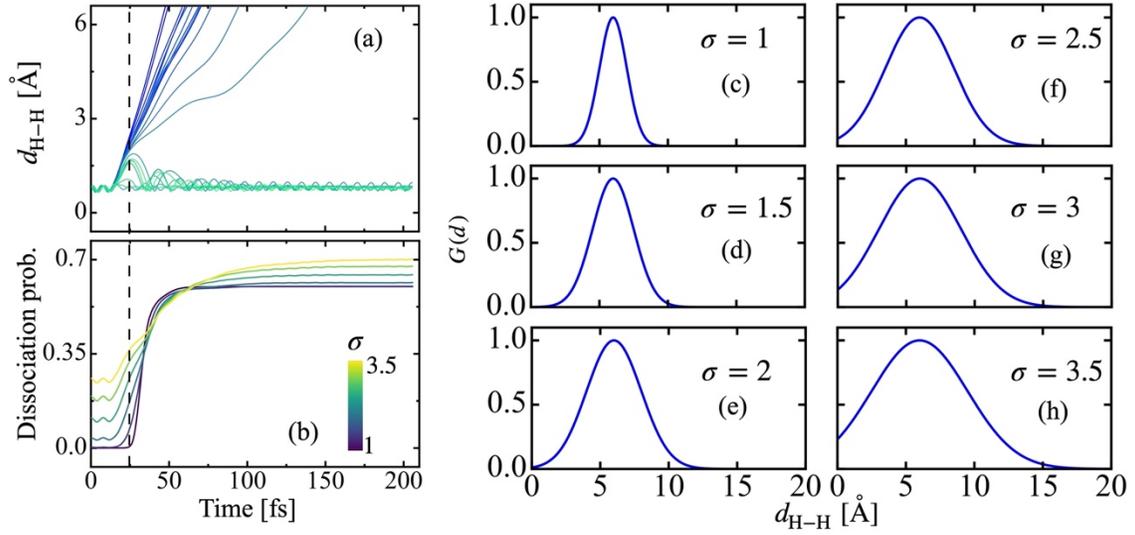

**Figure S3:** (a) Nuclear trajectories from Fig.3(d) in the main text and (b) dissociation probability as a function of time using a Gaussian weighting $G(d) = e^{-(d-d_0)^2/2\sigma^2}$ centered around $d_0 = 6$ Å and with widths ranging from $\sigma = 1$ to 3.5 Å. Note that as soon as $d$ passes 6 Å, the probability is kept at the peak value.